\documentclass{menu99}    
\usepackage{latexsym,graphicx,epsfig,psfrag}

\newcommand{\bea}{\begin{eqnarray}}
\newcommand{\eea}{\end{eqnarray}}

\newcommand{\fs}{\; .}
\newcommand{\co}{\; ,}

\newcommand{\eff}{{e\hspace{-0.1em}f\hspace{-0.18em}f}}
\newcommand{\mN}{m_{\hspace{-0.05em}\scriptscriptstyle N} }
\newcommand{\gpiN}{g_{\pi\hspace{-0.05em}{\scriptscriptstyle N}}}
\newcommand{\gA}{g_{\hspace{-0.05em}{\scriptscriptstyle A}}}
\begin{document}
    \setlength{\baselineskip}{2.6ex}

\title{Effective Field Theory of the Pion-Nucleon-Interaction\thanks{Work
    supported in part by Schweizerischer Nationalfonds}} 
\author{H. Leutwyler
\\{\em Institute for Theoretical Physics, University of Bern,
Sidlerstr. 5, CH-3012 Bern, Switzerland}}

\maketitle

\begin{abstract}
\setlength{\baselineskip}{2.6ex}
In the first part of the talk, I discussed the nature of the effective theory 
that describes the properties of the pion-nucleon-interaction at low energies,
using the static model as a starting point. In the second part, I then
pointed out that the infrared singularities occurring in the
$\pi$N scattering amplitude are stronger and of a more complex structure
than those encountered in $\pi\pi$ scattering. 
A formulation of the effective
theory that properly accounts for these was described and 
the results obtained
thereby were illustrated with a few examples. 
In the following, I restrict myself to a
discussion of some qualitative aspects of this work.
\end{abstract}

\setlength{\baselineskip}{2.6ex}

\section*{INTRODUCTION}

The static model represents a forerunner of the effective theories
of the pion-nucleon interaction used today. In this model, the
kinetic energy of the nucleon is neglected: The nucleon is described as
a fixed source that only carries spin and isospin
degrees of freedom.  For an excellent review of the
model and its application to several processes of interest, I refer to the
book of Henley and Thirring \cite{Henley Thirring}.

The systematic formulation of the effective theory relies on an expansion
of the effective Lagrangian in powers of derivatives and quark masses.
Chiral symmetry implies that the leading term of this expansion is fully
determined by the pion decay constant $F_\pi$ and by the nucleon matrix
element of the axial charge, $\gA$. Disregarding vertices with three or
more pions, the explicit expression for the leading term reads
\bea {\cal L}_{\eff}= -\frac{\gA}{2 F_\pi}\bar{\psi}\,\gamma^\mu\gamma_5
\,\partial_\mu
\pi\psi +\frac{1}{8 F_\pi^2}\bar{\psi}\,\gamma^\mu i[\pi,\partial_\mu\pi]\psi
+\ldots\eea
The success of the static model derives from the fact that it properly 
accounts for the first term on the right hand side -- in the
nonrelativistic limit, where the momentum of the nucleons is neglected
compared to the nucleon mass.

The static model is only a model. In order for the effective theory to
correctly describe the properties of QCD at low energies, that framework 
must be extended, accounting for the second term in the above
expression for the effective Lagrangian, for the vertices
that contain three or more pion fields, for the contributions
arising at higher orders of the derivative expansion, as well as for the
chiral symmetry breaking terms generated by the quark masses $m_u$, $m_d$. 
This can be done in a systematic manner, using a nonrelativistic expansion
for the nucleon kinematics. The resulting framework is called
``Heavy Baryon Chiral Perturbation
Theory'' (HBCHPT). It represents an extension of the static model that 
correctly accounts for nucleon recoil, order by order in the nonrelativistic
expansion (for reviews of this approach, see for instance ref.
\cite{HBCHPT}). 

As pointed out in ref. \cite{Becher 1}, the nonrelativistic expansion of the
infrared singularities
generated by pion exchange is a subtle matter. The HBCHPT
representations of the scattering amplitude or of the scalar nucleon form 
factor, for example, diverge in the vicinity of the point $t=4M_\pi^2$. 
The problem does not arise in
the Lorentz invariant approach proposed earlier \cite{Gasser Sainio Svarc}.
It originates in the fact that for some of the graphs, 
the loop integration cannot be interchanged 
with the nonrelativistic expansion. 

The reformulation of the effective theory given in ref. \cite{Becher 1} 
exploits the fact that the infrared singular part of the one loop integrals
can unambiguously be
separated from the remainder. To any finite order of
the nonrelativistic expansion, the regular part represents a polynomial in the 
momenta. Moreover, the singular and regular pieces 
separately obey the Ward identities of chiral symmetry. 
This ensures that a suitable renormalization of the effective coupling
constants removes the regular part altogether. The
resulting representation for the various quantities of interest combines the
virtues of the Heavy Baryon approach with those of the relativistic 
formulation of ref. \cite{Gasser Sainio Svarc}:
The perturbation series can be ordered
with the standard chiral power counting and manifest Lorentz invariance is
preserved at every stage of the calculation.
 
\section*{GOLDBERGER-TREIMAN RELATION}
As a first illustration of the method, I briefly discuss 
the relation between the pion-nucleon coupling constant and the axial charge
of the nucleon, obtained on the basis of a calculation of the $\pi$N scattering
amplitude to order $q^4$. A detailed account of this work is in preparation
\cite{Becher 2}. Throughout the following, I consider the 
isospin limit, $m_u=m_d$, and replace the quark masses by the leading term 
in the expansion of $M_\pi^2$,
\bea M^2\equiv (m_u+m_d) B\fs\nonumber\eea 
The Goldberger-Treiman relation may be written in the form
\bea \gpiN=\frac{\gA \mN}
{F_\pi}\{1+ \Delta_{\scriptscriptstyle GT}\}\fs\eea
If the quark masses $m_u$, $m_d$ are turned off, the strength of the $\pi N$
interaction is fully determined by $\gA$ and $F_\pi$: 
$\Delta_{\scriptscriptstyle GT}=0$. The effective theory allows us to analyze
the correction that arises for nonzero quark masses. The 
quantities $\gpiN$, $\gA$, $\mN$ and $F_\pi$ may be calculated in terms
of the effective coupling constants. The result takes the form of
an expansion in powers of $M$, i.e.~in powers of the quark masses.

Some of the graphs occurring within the effective theory
develop infrared singularities when the pion mass is sent to zero.
These manifest themselves through odd powers of $M$ and through logarithms
thereof. The expansion of the nucleon mass, for instance,
is of the form
\bea\label{mN} \mN=m+k_1 M^2 +k_2 M^3 +k_3 M^4 \ln \frac{M^2}{m^2} + k_4 M^4 +
O(M^5)\fs\eea
The first term, $m$, is the value of the nucleon mass in the chiral limit.
The coefficients $k_1,k_2,\ldots$ represent combinations of effective coupling
constants that remain finite when the quark masses are turned off.
In particular, the coefficient of the term proportional to $M^3$ is given by
\bea k_2=-\frac{3g^2}{32 \pi F^2}\co\eea
where $g$ and $F$ represent the values
of $\gA$ and $F_\pi$ in the chiral limit,
respectively. The term is correctly described by the static model, where
it arises from the self energy of the pion cloud that surrounds the nucleon.
Numerically, this contribution lowers the nucleon mass by about 15 MeV.

Similar terms also occur in the chiral
expansion of $\gA$ --  Kambor and  Moj\v zi\v s \cite{Kambor Mojzis} have
worked out this quantity to order $q^3$. The expansion of
$F_\pi$ is known since a long time -- it only contains even powers of $M$,
accompanied by logarithms. The coupling constant $\gpiN$ is obtained by
evaluating the residue of the 
pole terms occurring in the scattering amplitude at 
$s=\mN^2$ and $u=\mN^2$.
The representation of the amplitude to order $q^4$
yields an expression for $\gpiN$ in terms
of the effective coupling constants and of the quark masses, valid up to and
including order $q^3$. 

Using these results, we may evaluate the chiral expansion of
$\Delta_{\scriptscriptstyle GT}$ to order $q^3$. 
Remarkably, the contributions of order
$M^2\ln M^2/m^2$ as well as those of order $M^3$ cancel -- 
the Goldberger-Treiman
relation is free of infrared singularities,
up to and including order 
$M^3$:
\bea \Delta_{\scriptscriptstyle GT}=k_{\scriptscriptstyle
  GT}\,M^2+O(M^4)\fs\eea
The coefficient $k_{\scriptscriptstyle
  GT}$ represents a combination of effective coupling constants -- chiral
symmetry does not determine its magnitude.
The result shows that in the case of the Goldberger-Treiman relation,
the breaking of chiral symmetry generated by
the quark masses does not get enhanced by small energy denominators.  
Assuming that the scale of the symmetry breaking is the same as 
in the case of $F_K/F_\pi$, where it is
set by the massive scalar states, $M_S\simeq 1\,\mbox{GeV}$, we obtain the
crude estimate $\Delta_{\scriptscriptstyle GT}\simeq M_\pi^2/M_S^2\simeq 0.02$.
The detailed analysis based on models and on the SU(3) breaking effects
seen in the meson-baryon coupling constants \cite{Dominguez Gensini Thomas} 
confirms the expectation that $\Delta_{\scriptscriptstyle GT}$ must be
very small -- a discrepancy of order 4\% or more would be very difficult to
understand.

Since the days when the Goldberger-Treiman relation was discovered,
the value of $\gA$ has increased considerably. Also, 
$F_\pi$ decreased a little, on account of radiative corrections. The main
source of uncertainty is $\gpiN$. The comprehensive analysis of $\pi N$
scattering published by
H\"ohler in 1983 \cite{Hoehler} led to 
$f^2=\gpiN^2 M_\pi^2/(16 \pi \mN^2)=0.079$. 
With $\gA=1.267$ and $F_\pi=92.4 \,\mbox{MeV}$, this value yields
$\Delta_{\scriptscriptstyle GT}=0.041$.
As stressed by Pavan at this meeting, the data accumulated since then
indicate that $f^2$ is somewhat smaller, 
numbers in the range from 0.076 to 0.077 looking more likely. This range
corresponds to $0.021<\Delta_{\scriptscriptstyle GT}<0.028$. 

I conclude that, within the current experimental uncertainties to be attached
to the pion-nucleon coupling constant, the Goldberger-Treiman relation does
hold to the expected accuracy. Note that at the level of 1 or 2 \%,
isospin breaking cannot be ignored. In particular, 
radiative corrections need to be analyzed carefully. Also, the coupling
constant relevant for the neutral pion picks up a significant contribution
from $\pi^0-\eta$ interference. A precise determination of the pion-nucleon
coupling constant is essential to arrive at reliable results for small
quantities such as the $\sigma$-term. The various discussions  
at this meeting show that the issue is under close scrutiny by
several groups and I am confident that the uncertainties will soon be reduced. 

\section*{LOW ENERGY THEOREM FOR ${\bf D^+}$}
As a second example, I consider the low energy theorem for the value of the
scattering amplitude $D^+(s,t,u)$ at the Cheng-Dashen point:
$s=u=\mN^2$, $t= 2 M_\pi^2$. The theorem relates this amplitude to the scalar 
form factor of the nucleon,
\bea \langle N'|\,m_u \,\bar{u}u+m_d\, \bar{d}d\,|N\rangle=
\sigma(t)\bar{u}' u\fs\eea
The relation may be written in the form
\bea\label{CD} F_\pi^2 D^+(\mN^2,2M_\pi^2,\mN^2)=
\sigma(2M_\pi^2)+\Delta_{\scriptscriptstyle CD}\fs\eea
The theorem states that the term $\Delta_{\scriptscriptstyle CD}$ 
vanishes up to and including contributions of order $M^2$. 
The explicit expression
obtained for $F^2_\pi D^+(\mN^2,2M_\pi^2,\mN^2)$ when evaluating the
scattering amplitude to order $q^4$ again contains infrared singularities
proportional to $M^3$ and $M^4\ln M^2/m^2$. Precisely the same singularities,
however, also show up in the scalar form factor at $t=2M_\pi^2$, 
so that the result for
$\Delta_{\scriptscriptstyle CD}$ is free of such singularities\footnote{The
  cancellation of the terms of order $M^3$ was pointed out in
  ref. \cite{Gasser Sainio Svarc,Pagels Pardee} and the absence of logarithmic
contributions of order $M^4$ was shown in ref. \cite{BKM}.}
\bea \Delta_{\scriptscriptstyle CD}=k_{\scriptscriptstyle CD}M^4+O(M^5)
\fs\eea
Crude estimates like those used in the case of the Goldberger-Treiman
relation indicate that the term 
$\Delta_{\scriptscriptstyle CD}$ must be very small, of order 1 MeV.

The value of the scalar form factor at $t=0$ is referred to as the
$\sigma$-term,
\bea \sigma=\sigma(0)\fs\nonumber\eea
This quantity is of particular interest, because it
represents the response of the nucleon mass to a change in the quark masses:
\bea\label{FH} \sigma=m_u\frac{\partial \mN}{\partial m_u}+
m_d\frac{\partial \mN}{\partial m_d}\fs\eea
The difference $\sigma(2M_\pi^2)-\sigma(0)$ is well understood: The
results found on the basis of a dispersive analysis \cite{Gasser Leutwyler
  Sainio} are confirmed by
the evaluation of the scalar form factor within the effective
theory \cite{BKM 93,Becher 1}. 
The net result is that an accurate determination of the
scattering amplitude at the Cheng-Dashen point amounts to 
an accurate determination of the $\sigma$-term.

Unfortunately, the experimental situation concerning the magnitude
of $D^+$ at the Cheng-Dashen point leaves much to be desired (for a recent
review, see ref. \cite{Sainio}). The low energy
theorem makes it evident 
that we are dealing with a small quantity here -- the object
vanishes in the chiral limit. The
inconcistencies among the various data sets available at low energies
need to be clarified to arrive at a reliable value for $\gpiN$. Only then
will it become possible to accurately measure 
small quantities such as the $\sigma$-term.

\section*{DEPENDENCE OF THE ${\bf \sigma}$-TERM ON THE QUARK MASSES} 
In the following, I do not discuss the magnitude 
of $\sigma$ as such, but instead consider the dependence of this quantity on 
the quark masses, which is quite remarkable. 
In this discussion, the precise value of
$\sigma$ is not of crucial importance. For definiteness, I use the value
$\sigma=45\,\mbox{MeV}$ \cite{Gasser Leutwyler
  Sainio}. The chiral expansion of the $\sigma$-term is readily obtained by
applying the Feynman-Hellman theorem (\ref{FH}) to 
the formula (\ref{mN}), with the result
\bea\label{sigma} \sigma=k_1 M^2 +\frac{3}{2}\,k_2 M^3 +k_3 M^4 
\left\{2\,\ln \frac{M^2}{m^2}+1\right\} + 
2\, k_4 M^4 +
O(M^5)\co\eea 
As discussed above, the term proportional to $M^3$ arises from an infrared
singularity in the self energy of the pion cloud. It lowers the magnitude of
the $\sigma$-term
by $3/2\times 15 \,\mbox{MeV} \simeq 23\, \mbox{MeV}$. 
The coefficient $k_3$ can also be expressed in terms
of measurable quantities \cite{Becher 1}. Numerically the contribution from
this term amounts
to $- 7 \,\mbox{MeV}$, thus amplifying the effect seen at $O(M^3)$. 
Chiral symmetry does not determine all of 
the effective coupling constants entering the regular contribution 
$k_4 M^4$, which is of the same type as the correction 
$\Delta_{\scriptscriptstyle CD}= 
k_{\scriptscriptstyle CD}M^4$ to the low energy
theorem (\ref{CD}). As discussed above, corrections of this type are
expected to be very small -- I simply drop the term $k_4 M^4$.
The value of $k_1$ is then fixed by the input $\sigma=45\,\mbox{MeV}$
for the total,  so that we can now discuss the manner in which $\sigma$
changes when the quark masses are varied. 

\begin{figure}

\leavevmode   
\psfrag{x}{$m_u+m_d$}
\psfrag{y}{$\sigma$}
\psfrag{z}{\raisebox{-0.3em}{$\uparrow$}}

\centering

\includegraphics[width=8cm]{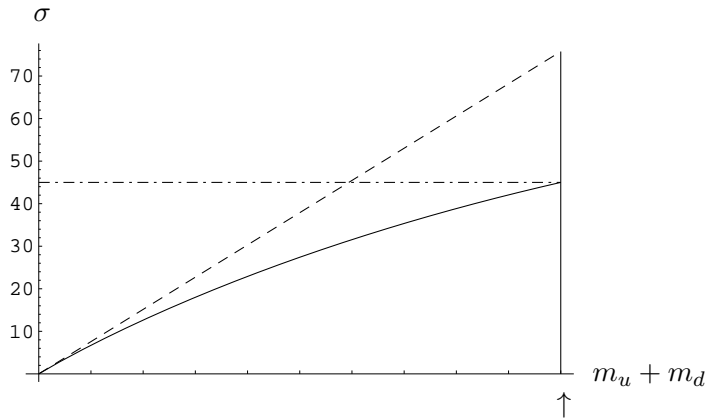}
\caption{ $\sigma$-term (in MeV) as a function of the quark
  masses. It is assumed that the physical value of $\sigma$ is 45 MeV 
(dash-dotted line).
The arrow corresponds to the physical value of $m_u+m_d$. 
 The dashed line depicts the linear dependence that
results if the infrared singular 
contributions proportional to $M_\pi^3$ and to $M_\pi^4 \ln
M_\pi^2/\mN^2$ are dropped. }
\end{figure}

At leading order, $\sigma$ is given by  $k_1 M^2$. In figure 1 
this contribution is shown as a dashed straight line. The full curve 
includes the contributions generated by the infrared singularities,
$k_2 M^3$ and $k_3 M^4\{2 \ln M^2/m^2+1\}$. 
The figure shows that the expansion of the 
$\sigma$-term in powers of the quark masses contains large
contributions from infrared singularities. These must show up in evaluations
of the $\sigma$-term on a lattice: The ratio $\sigma/(m_u+m_d)$ must change
significantly if the quark masses are varied from the chiral limit to their
physical values. Note that in this discussion, 
the mass of the strange quark is kept fixed at
its physical value -- the curvature seen in the figure exclusively arises
from the perturbations generated by the two lightest quark masses.

It is instructive to compare this result with the dependence of $M_\pi^2$ on
the quark masses. In that case, the expansion only contains even powers of
$M$:
\bea M_\pi^2 =M^2 +\frac{M^4}{32 \pi^2 F^2}\ln\frac{M^2}{\Lambda_3^2} + 
O(M^6)\fs\eea
The quantity $\Lambda_3$ stands for the renormalization group invariant scale
of the effective coupling constant $l_3$. The SU(3) estimate for this
coupling constant given in ref. \cite{GL 1984} reads $\bar{l}_3\equiv -\ln
M_\pi^2/\Lambda_3^2 =2.9\pm 2.4$. The error bar is so large that the estimate 
barely determines the order of magnitude 
of the scale $\Lambda_3$.
Figure 2 shows, however, that this uncertainty does not significantly affect 
the dependence of $M_\pi^2$ on the quark masses, because the logarithmic
contribution is tiny: the range of $\bar{l}_3$ just quoted
corresponds to the shaded region shown in the figure.

\begin{figure}

\leavevmode   
\psfrag{x}{$m_u+m_d$}
\psfrag{y}{$M_\pi^2$}
\centering

\includegraphics[width=8cm]{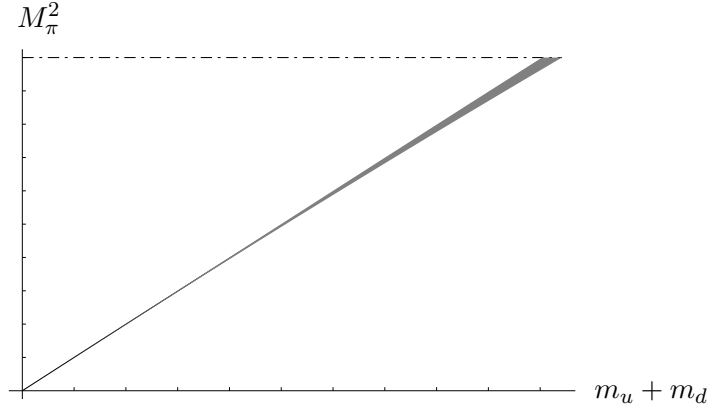}

\caption{Square of the pion mass as a function of the quark masses.
The
dash-dotted line indicates the physical value of $M_\pi^2$.}
\end{figure}

The logarithmic term occurring in the chiral expansion of $M_\pi^2$
gets enhanced by about a factor of two if
we consider the pion $\sigma$-term,  
\bea\sigma_\pi=\langle\pi|\,
m_u\,\bar{u}u+m_d\,\bar{d}d\,|\pi\rangle=
m_u\frac{\partial M_\pi^2}{\partial m_u}+ m_d
\frac{\partial M_\pi^2}{\partial m_d}\fs \eea 
I do not show the corresponding curve, because it is also nearly a straight
line. 

The main point here is that
the infrared singularities encountered in the self energy of the 
nucleon are much stronger than those occurring in the self energy of the pion.
A detailed account of the work on the low energy structure of the
$\pi N$ scattering amplitude done in collaboration with Thomas Becher
is in preparation.
 
\bibliographystyle{unsrt}

\begin{thebibliography}{99}
\bibitem{Henley Thirring} E.M.~Henley and W.~Thirring, {\em Elementary Quantum
  Field Theory}, (McGraw-Hill, New York 1962), Part III. Pion Physics.
\bibitem{HBCHPT} G.~Ecker,  Czech.\ J.\ Phys. {\bf 44}, 505 (1994).\\ 
V.~Bernard, N.~Kaiser and U.-G.~Meissner, Int.\ J.\ Mod.\ Phys. {\bf E4}, 193 
(1995).
\bibitem{Becher 1} T.~Becher and H.~Leutwyler,
Eur.\ Phys.\ J. {\bf C9} 643 (1999). For earlier work in this direction see
P.~J.~Ellis and H.-B.~Tang, Phys.\ Rev.\ {\bf C57} 3356 (1998).
\bibitem{Gasser Sainio Svarc}
  J.~Gasser, M.~E.~Sainio and A.~Svarc, Nucl.\ Phys.\ {\bf B307} 779 (1988).
\bibitem{Becher 2} T.~Becher and H.~Leutwyler, to be published.
\bibitem{Kambor Mojzis} J.~Kambor and M.~Moj\v zi\v s, 
``Field redefinitions 
and wave function 
renormalization to $O(q^4)$ in HBCHPT'', J.\ High\ Energy\ Physics {\bf 9904},
  031(1999)    
\bibitem{Dominguez Gensini Thomas}
C.A.~Dominguez, Phys.Rev. {\bf D27} 1572 (1983); Nuovo Cim. {\bf 8}, 1
  (1985).\\ 
P.A.M.~Guichon, G.A.~Miller and A.W.~Thomas, Phys.Lett. {\bf 124B} 109 
(1983).\\
P.M.~Gensini, Nuovo Cim. {\bf 102A}, 75 (1989), E: ibid. {\bf 102A} 1181
  (1989). 
\bibitem{Hoehler}
G.~H\"ohler, in Landolt-B\"ornstein, {\bf 9b2}, 
ed.~H.~Schopper (Springer, Berlin, 1983).

\bibitem{Pagels Pardee}H.~Pagels and W.J.~Pardee, Phys.Rev. {\bf D4}, 3335
  (1971). 

\bibitem{BKM}V.~Bernard, N.~Kaiser and U.~Meissner, Phys.Lett. {\bf B389},
144 (1996).  
\bibitem{Gasser Leutwyler Sainio}
J.~Gasser, H.~Leutwyler and M.~E.~Sainio, Phys.\ Lett. {\bf B253},
  252, 260  (1991). 
\bibitem{BKM 93}V.~Bernard, N.~Kaiser and U.~Meissner, Z.Phys. {\bf C60}, 111
  (1993).  
\bibitem{Sainio}M.E.~Sainio, ``Low-energy pion-nucleon interaction and the
  sigma term'',\\ $\pi N$ Newslett. {\bf 13} 144 (1997).

\bibitem{GL 1984}J. Gasser and H. Leutwyler, Ann.\ Phys.\ {\bf 158} 142 (1984).

\end{thebibliography}

\end{document}